\documentclass[doublecol]{epl2} 
\usepackage{amsmath}
\usepackage[T1]{fontenc}

\newcommand\hbtd{{\sf{HB-T}}}
\newcommand\hbt{{\sf{HB-T}~}}

\newcommand{\beq}{\begin{equation}}
\newcommand{\eeq}{\end{equation}}
\newcommand{\bea}{\begin{eqnarray}}
\newcommand{\eea}{\end{eqnarray}}

\def\ket#1{| \,#1\, \rangle}

\def\expect#1{\langle \,#1\, \rangle}

\def\expect#1{\langle \,#1\, \rangle}

\title{Non-local geometric phase in two-photon interferometry}
\author{A. Martin\inst{1} \and O. Alibart\inst{1} \and J.-C. Flesch\inst{1} \and J. Samuel\inst{2} \and Supurna Sinha\inst{2} \and S. Tanzilli\inst{1} \and A. Kastberg\inst{1,2}}
\institute{                    
  \inst{1} Laboratoire de Physique de la Matière Condensée - CNRS UMR 6622, Université de Nice -- Sophia Antipolis, Parc Valrose, Nice Cedex 2, France\\
  \inst{2} Raman Research Institute - Bangalore 560 080, India
}

\abstract{
We report the experimental observation of the nonlocal geometric phase in Hanbury Brown-Twiss polarized intensity interferometry. The experiment involves two independent, polarized, incoherent sources, illuminating two polarized detectors. Varying the relative polarization angle between the detectors introduces a geometric phase equal to half the solid angle on the Poincar\'e sphere traced out by a \emph{pair} of single photons. Local measurements at either detector do not reveal the effect of the geometric phase, which appears only in the coincidence counts between the two detectors, showing a genuinely nonlocal effect. We show experimentally that coincidence rates of photon arrival times at separated detectors can be controlled by the two photon geometric phase. This effect can be used for manipulating and controlling photonic entanglement.}

\begin{document}

\maketitle

\section{Introduction}
\label{intro}
Quantum interference is one of the most delicate and intriguing aspects of quantum theory. 
A subtle interference effect, discovered surprisingly late, is due to the geometric phase~\cite{Berry:1984,Shapere:1989}. It was soon realized~\cite{Ramaseshan:1986} that Berry's discovery had been anticipated by Pancharatnam's work on the interference of polarized light~\cite{Pancharatnam:1956}. Pancharatnam's work is now widely recognized 
as an early precursor and optical analog of the geometric phase~\cite{Ramaseshan:1986,Samuel:1988}. 

\begin{figure}%[h]
\onefigure[width=0.9\columnwidth]{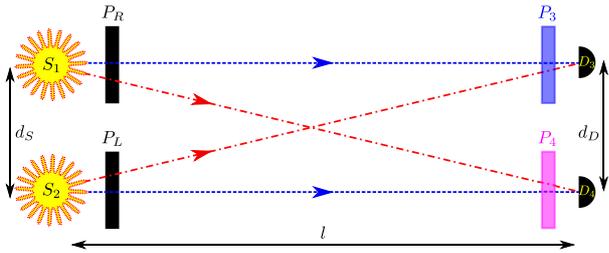}
\caption{
General scheme for the Pancharatnam phase dependent two-photon intensity interferometry. Two sources illuminate two detectors by direct processes (blue, dotted lines), and by exchange processes (red, dashed-dotted lines). The sources are respectively followed by analyzers for opposite circular polarization, and analyzers for different linear polarizations are placed in front of the detectors.}
\label{fig1}
\end{figure}

Studies of the geometric phase usually deal with \emph{amplitude} interferometry, which describes the interference of particles from coherent sources. Of particular interest to us here are multiparticle interference effects, discovered by Hanbury Brown and Twiss (\hbtd), who performed \emph{intensity} interferometry experiments using incoherent thermal sources~\cite{HanburyBrown:1956}. In a previous paper~\cite{Mehta:2010}, an experiment was proposed to detect a nonlocal Pancharatnam phase in polarized intensity interferometry. The idea of the experiment, \emph{c.f.} Fig.~\ref{fig1}, is to have two thermal sources of opposite circular polarizations, $P_\textrm{R}$ and $P_\textrm{L}$, where R and L describe right and left hand circular polarizations respectively. These illuminate two detectors that are covered by linear polarization analyzers, $P_\textrm{3}$ and $P_\textrm{4}$. A theoretical analysis of the experiment reveals that as the relative angle between the two linear polarizations, $\varphi_{34}$, is changed, the rate of coincidence counts between the arrival time of photons at the two detectors varies sinusoidally~\cite{Mehta:2010}.

In recent years there have  been many interferometric studies involving pairs of photons \cite{Klyshko:1990,Klyshko:1989,Shih:2011,Brendel:1999,Hariharan:2003}. Most of these studies use entangled photon pairs generated by spontaneous parametric down conversion and address effects like teleportation and Bell's inequality. There has been one study of geometric phases in two-photon interference~\cite{Brendel:1999} involving parametrically down converted photon pairs focussing on the interplay between geometric and  dynamic phases. In contrast, in our work work we use thermal sources to generate photon pairs and study the role of the geometric phase in inducing non-local cross correlations.

%%%%%%%%%%%%%%%%%%%%%%%%%%%%%%%%%%%%%%%
%%%%%%%%%%%%%%%%%%%%%%%%%%%%%%%%%%%%%%%
%%%%%%%%%%%%%%%%%%%%%%%%%%%%%%%%%%%%%%%
%%%%%%%%%%%%%%%%%%%%%%%%%%%%%%%%%%%%%%%
Entanglement lies at the very heart of quantum physics. It is now widely recognized as a fundamental resource in the growing field of quantum information, regarding both quantum computation and communication protocols~\cite{Tittel:2001,Nielsen:2010}. As with any resource, effective use requires control, and quantum information applications are especially concerned with the degree of entanglement between spatially separated quantum channels or systems. A multiparticle Aharonov-Bohm effect was discussed theoretically by B{\"u}ttiker~\cite{Buttiker:1992} who noted in the context of electronic charge transport that two-particle correlations can be sensitive to a magnetic flux even if the single particle observables are flux insensitive. The effect of the flux is visible only in current cross correlations and is a genuinely nonlocal and multiparticle Aharonov-Bohm effect~\cite{Aharonov:1959}. This has been experimentally seen in intensity interferometry experiments carried out using edge currents in quantum Hall systems~\cite{Neder:2007}, and the theory is further developed in~\cite{Samuelsson:2004,Splettstoesser:2009}. These authors propose an experiment in which Aharonov-Bohm fluxes may be used to control the orbital entanglement of electronic states. While this proposal is sound in principle, in practice, such experiments may be afflicted by decoherence~\cite{Nielsen:2010,Tittel:2001}. In contrast to electronic states, photonic states are easier to generate, manipulate and distribute. Most studies related to quantum communication protocols use photons rather than massive particles. However, photons are neutral and unaffected by the magnetic fields needed in the Aharonov Bohm effect. The idea here is to use the geometric phase as an effective magnetic field to control the degree of photonic entanglement~\cite{Mehta:2010}.

As a step towards achieving this goal, we report in this letter an experiment based on two independent polarized thermal sources for a demonstration of the nonlocal Pancharatnam phase effect. Such a nonlocal effect can only be observed in the intensity cross correlations ${\cal G}^{(2)}$. A previously reported experiment involving single photons prepared in mixed polarization states led to lower order correlations ${\cal G}^{(1)}$~\cite{Ericsson:2005}. Here we take advantage of the setup suggested in~\cite{Mehta:2010} and show that the geometric phase effect only appears in the coincidence counts between spatially separated detectors when the relative angle between the polarization analysers is tuned. The two photon Pancharatnam phase is nonlocal in the precise sense that it cannot be observed when local measurements at either detector are performed. The coincidence counts between the two detectors are modulated by a phase which has a geometric component as well as a dynamical (or propagation) phase. 

Ideally, the experiment should use \emph{a priori} incoherent thermal sources, as in the \hbt experiment. However, natural thermal light sources would necessitate very short integration times (shorter than the coherence time), which is hard to implement experimentally. Standard laser sources, which are usually used in amplitude interferometry, are unsuitable since they are coherent and do not show the \hbt effect. In this experiment, we use a laser source and introduce incoherence by passing the laser beam through a ground glass plate~\cite{Arecchi:1965}. This gives us a source of light in which the amplitude and phase are randomly varying, mimicking a thermal light source intensity distribution. Using such a (pseudo-thermal) source, we demonstrate the nonlocal Pancharatnam phase~\cite{Mehta:2010}.

\section{Theory}
\label{theory}
The central quantity of interest is the polarized intensity-intensity correlation function measured by the coincidence counts for the detectors D$_3$ and D$_4$, given by: 
%%%%%%%%%%%%%%%%%%%%%%%%%%%%%%%%%%%%%%%
\begin{equation}
{\cal C} = {\cal G}^{(2)}_{34} = \dfrac{\expect{N_3 N_4}}{\expect{N_3}\expect{N_4}} \, , 
\label{equ1}   
\end{equation}
%%%%%%%%%%%%%%%%%%%%%%%%%%%%%%%%%%%%%%%
$N_3$ and $N_4$ being the photon numbers detected at D$_3$ and D$_4$ per time per bandwidth. The detailed theory is outlined in~\cite{Mehta:2010}. The theoretical prediction for ${\cal C}$ is:
%%%%%%%%%%%%%%%%%%%%%%%%%%%%%%%%%%%%%%%
\begin{equation}
{\cal C} = \frac{3}{2} + \frac{1}{2} \cos \left[ k[(r_{13} -r_{14})-(r_{23} -r_{24})]+\frac{\Omega}{2}\right]
\label{equ2} \, ,  
\end{equation}
%%%%%%%%%%%%%%%%%%%%%%%%%%%%%%%%%%%%%%%
where $r_{ij}$, with $i=1,2$ and $j=3,4$, denote the optical path lengths between the sources $S_1$ and $S_2$ and detectors $D_3$ and $D_4$. $\Omega$ is the solid angle enclosed by the geodesic path shown in Fig. $2$.

\begin{figure}%[h]
\onefigure[width=0.5\columnwidth]{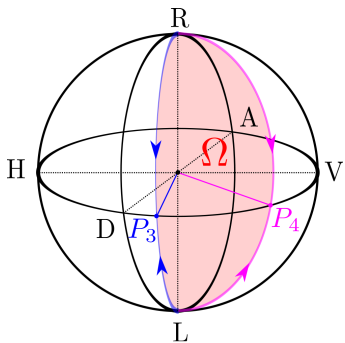}
\caption{
\label{fig2}
Path on the Poincaré sphere determining the detected geometric phase. The relative angle $\varphi_{34}$ between the linear polarizers $P_3$ and $P_4$ sets the angular width of the enclosed surface on the sphere. $\Omega=4\varphi_{34}$ corresponds to the solid angle defined by the geodesic path R~$\rightarrow P_3 \rightarrow$~L~$\rightarrow P_4 \rightarrow$~R.}
\end{figure}

This theoretical calculation can be summarized as follows. There is a process in which a photon from S$_1$ reaches detector D$_3$ and a photon from S$_2$ reaches D$_4$ (direct process, shown in blue). There is also a corresponding exchange process in which a photon from S$_1$ reaches D$_4$ and one from S$_2$ reaches D$_3$ (exchange process, shown in red). These two processes are indistinguishable, and in the absence of any `welcherweg' information, we have to combine the amplitudes for these processes with a relative phase $\phi$, which depends on the optical path difference between the two processes. The quantum interference between these processes causes the \hbt effect. Only the interference between direct and exchange processes gives rise to oscillations; the remaining terms providing the background of 1 in ${\cal G}^{(2)}$. We expect the coincidence counts to vary with the propagation phases and thus, the counts should depend on the detector separation $d_\mathrm{D}$ and the wavelength $\lambda$ of the light. The new effect that is present in this polarized version of \hbt is that the coincidence counts also depend on the difference between the linear polarization angles for $P_3$ and $P_4$, $\varphi_{34}$, and is modulated by a geometric phase of half the solid angle ${\Omega}$ covered on the Poincar\'e sphere by the four polarizations involved (see Fig.~\ref{fig2}). The closed path on the Poincar\'e sphere is a property of a \emph{pair} of photons. The effect of the geometric phase on the coincidence rates of photons at the detectors D$_3$ and D$_4$ is what we measure in the experiment below.

%%%%%%%%%%%%%%%%%%%%%%%%%%%%%%%%%%%%%%%
%%%%%%%%%%%%%%%%%%%%%%%%%%%%%%%%%%%%%%%
%%%%%%%%%%%%%%%%%%%%%%%%%%%%%%%%%%%%%%%
%%%%%%%%%%%%%%%%%%%%%%%%%%%%%%%%%%%%%%%

\section{Experiment}
\label{experiment}
\begin{figure}%[h]
\onefigure[width=85mm]{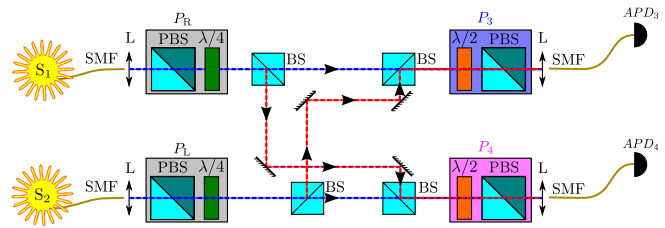}
\caption{
\label{fig3}
Experimental setup for demonstrating the Pancharatnam phase at the two-photon level. S$_1$ and S$_2$ : independent pseudo-thermal light sources; SMF: single-mode optical fibre; L: lens; PBS: polarizing beamsplitter; $\lambda/4$ and $\lambda/2$: quarter- and half-waveplates, respectively; BS: non-polarizing beamsplitter; APD: avalanche photodiode.}
\end{figure}

Our experimental realization of Fig.~{\ref{fig1}} is schematically shown in Fig.~{\ref{fig3}}. Two independent pseudo-thermal light sources (S$_1$ and S$_2$) are spatially filtered using single mode fibers, and their polarizations are adjusted to right-hand and left-hand circular ($P_\textrm{R}$ and $P_\textrm{L}$ for S$_1$ and S$_2$, respectively) using polarizing beamsplitters (PBS) and quarter-wave plates.
%%%%%%%%%%%%%%%%%%%%%%%%%%%%%%%%%%%%%%%
%%%%%%%%%%%%%%%%%%%%%%%%%%%%%%%%%%%%%%%
%%%%%%%%%%%%%%%%%%%%%%%%%%%%%%%%%%%%%%%
%%%%%%%%%%%%%%%%%%%%%%%%%%%%%%%%%%%%%%%
The output photons from the sources are first directed towards a set of non-polarizing beamsplitters (BS) in order to split each source into two paths. The four resulting paths are led towards the linear polarization analyzers $P_3$ and $P_4$. A second set of BS is then used to recombine, before each analyzer, two paths, with identical probability amplitudes, emanating from S$_1$ and S$_2$. 
The linear analyzers are each a half-waveplate and a PBS, and the detectors are two independent silicon avalanche photodiodes (APD), featuring 5\% detection efficiencies and dark count rates on the order of 100/s.

Our pseudo-thermal sources emit monochromatic, thermal light, with an adjustable coherence time. This is accomplished by focusing a 852~nm external cavity diode laser on a rotating, ground glass disk, producing time dependent fluctuations in the beam~\cite{Martienssen:1964,Scarcelli:2004}. The scattered light is then collected using two single-mode optical fibers, which provide suitable spatial mode selections. The light arrives at the observation point, P, scattered through different routes. The electric field intensity received can be written as a sum over scatterers $E=\Sigma_i E_i\exp{(\mathrm{i}\phi_i)}$, where the phase $\phi_i$ and the amplitude $E_i$ due to the $i^{\mathrm{th}}$ scatterer are both random. The statistics of the intensity fluctuations is given by a Rayleigh distribution~\cite{Arecchi:1965,Loudon:2000}
%%%%%%%%%%%%%%%%%%%%%%%%%%%%%%%%%%%%%%%
\begin{equation}
P(I)=\frac{1}{I_0} \exp{\left( -\frac{I}{I_0} \right) } \, .
\label{equ3}
\end{equation}
%%%%%%%%%%%%%%%%%%%%%%%%%%%%%%%%%%%%%%%
The light received at P becomes uncorrelated with itself as the the beam traverses the plate. For a Gaussian beam profile incident on the glass plate, we expect the intensity autocorrelation function ${\cal G}^{(2)}(\tau) = \langle I(0)I(\tau) \rangle / (\langle I(0) \rangle \langle I(0) \rangle)$ to show a Gaussian decay in time~\cite{Loudon:2000}. More precisely, ${\cal G}^{(2)}(\tau)= 1+\exp{(-\pi(\tau/\tau_\mathrm{c})^2)}$, where the coherence time $\tau_\mathrm{c}$ is on the order of the time taken by a point on the glass plate to traverse the beam width. Such a Gaussian decay of correlations is indeed observed in our experiment.

\begin{figure}%[]
\onefigure[width=\columnwidth , height=55mm]{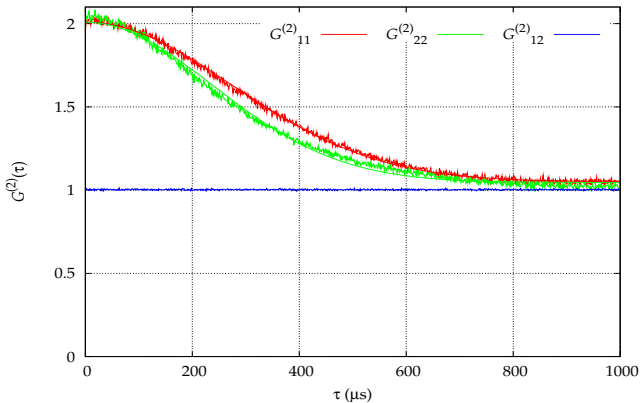}
\caption{
\label{fig4}
Autocorrelation functions (${\cal G}^{(2)}_{11}(\tau)$ and ${\cal G}^{(2)}_{22}(\tau)$) of the pseudo-thermal light sources S$_1$ and S$_2$ (red and green), and their crosscorrelation function (${\cal G}^{(2)}_{12}(\tau)$ -- blue), as functions of time using sampling frequency of 1~MHz.}
\end{figure}

The average intensities have been chosen such that both the probability of having more than one photon during the dead-time of the APDs (45~ns), and the probability of having a dark count, are negligible. The effective coherence times of the sources, and the absence of correlations between them, have been tested by measuring their autocorrelations (${\cal G}^{(2)}_{ii}(\tau)$) and the crosscorrelation function (${\cal G}^{(2)}_{ij}(\tau)$). The corresponding experimental results are shown in Fig.~{\ref{fig4}}. 
%%%%%%%%%%%%%%%%%%%%%%%%%%%%%%%%%%%%%%%
%%%%%%%%%%%%%%%%%%%%%%%%%%%%%%%%%%%%%%%
%%%%%%%%%%%%%%%%%%%%%%%%%%%%%%%%%%%%%%%
%%%%%%%%%%%%%%%%%%%%%%%%%%%%%%%%%%%%%%%
As predicted, both sources feature a Gaussian correlation decay as a function of time, associated with coherence times of about 7-800~$\mu$s. In addition they exhibit no crosscorrelation, proving they are actually independent. Therefore, we have estimated that sampling frequencies down to 100~kHz will ensure good measurements of ${\cal G}^{(2)}(0)$, while $10^5$ samples give us a negligible error. These parameters have been chosen in order to minimize the measurement time, and thus avoiding any unwanted path length fluctuations in the interferometer, as well as frequency drifts of the laser.

To observe the interference pattern in the coincidence counts ${\cal C}$ (see Eq.~{\ref{equ2}), we rotate the $P_3$ analyzer's half-waveplate by an angle $\theta$ from 0 to $2\pi$, thereby continuously scanning the polarization (and thus the angle $\varphi_{34}$, from 0 to $4\pi$). The recorded signal is shown in Fig.~{\ref{fig5}}, as well as the expected fringe pattern with an ideal visibility.
%%%%%%%%%%%%%%%%%%%%%%%%%%%%%%%%%%%%%%%
%%%%%%%%%%%%%%%%%%%%%%%%%%%%%%%%%%%%%%%
\begin{figure}%[h]
\onefigure[width=\columnwidth]{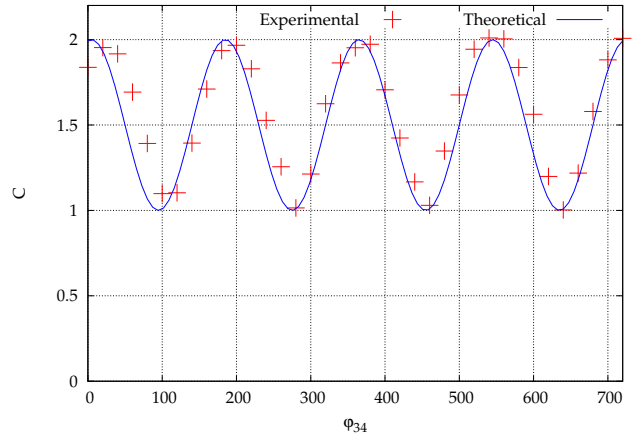}
\caption{
\label{fig5}
Crosscorrelation function ${\cal C} = {\cal G}^{(2)}_{34}(0)$, as function of the angle $\varphi_{34}=2\theta$ between the polarizations $P_3$ and $P_4$.}
\end{figure}
%%%%%%%%%%%%%%%%%%%%%%%%%%%%%%%%%%%%%%%
%%%%%%%%%%%%%%%%%%%%%%%%%%%%%%%%%%%%%%%

\section{Discussion}
\label{discussion}
Apart from its intrinsic interest as a demonstration of the nonlocal geometric phase in optics, one of our prime motivations is to realize a controllable and reconfigurable circuit enabling the generation and the manipulation of all the four photonic entangled Bell states~\cite{Nielsen:2010}. This can be done by extending our experiment in the following way. Let us first replace the two thermal sources of Fig.~\ref{fig1} by two simultaneously heralded single photon sources. We suppose that when two photons are emitted at the same time, \emph{i.e.}, one per source, they have opposite circular polarization. Second, we interface these polarized single photon emitters and the current setup by a time-bin preparation circuit acting on the two photons~\cite{Tittel:2001}, where time-bin here is the means for creating coherent superpositions of states in the time domain. This way, using beamsplitters and identical delay lines for the two photons, we can make these photons travel over short, $\ket{\mathrm{s}}$, or long, $\ket{\mathrm{l}}$, paths, with certain probability amplitudes, and accordingly prepare time-bin input qubits for the geometric phase setup. We prepare incident photons initially with $P_\mathrm{R}$ and $P_\mathrm{L}$ polarization in the time-bin superpositions $\ket{\varphi} = \alpha \ket{\mathrm{s}} + \beta\ket{\mathrm{l}}$ and $\ket{\psi} = \alpha' \ket{\mathrm{s}} + \beta'\ket{\mathrm{l}}$, respectively. Here, $\alpha$, $\beta$, $\alpha'$, and $\beta'$ are complex numbers related to the beamsplitting ratios. In this configuration~\cite{Scarcelli:2004}, the output state is given by:
%%%%%%%%%%%%%%%%%%%%%%%%%%%%%%%%%%%%%%%
\begin{equation}
\ket{\Upsilon} = \frac{1}{\sqrt{2}}\left(\ket{\varphi_3}\ket{\psi_4} + \textrm{e}^{\textrm{i}\phi}\ket{\psi_3}\ket{\varphi_4}\right) \, ,
\label{equ4}
\end{equation}
%%%%%%%%%%%%%%%%%%%%%%%%%%%%%%%%%%%%%%%
where $\phi$ can be tuned using the geometric phase.

On one hand, simply setting the input states to $\ket{\varphi}=\ket{\mathrm{s}}$ and $\ket{\psi}=\ket{\mathrm{l}}$ permits preparing time-bin entangled states of the form $(\ket{\mathrm{sl}} + \mathrm{e}^{\textrm{i}\phi}\ket{\textrm{ls}})/\sqrt{2}$. Such states are otherwise non trivial to produce by usual experimental means based on pulsed parametric down-conversion associated with a preparation interferometer~\cite{Brendel:1999}. Here, the geometric phase allows switching between the $\ket{\Psi^+}$ and the $\ket{\Psi^-}$ Bell states.

On the other hand, substituting in Eq.~\ref{equ4} superposition input states of the form $\ket{\varphi} = (\ket{\mathrm{s}}+\textrm{e}^{\textrm{i}\phi_1}\ket{\mathrm{l}})\sqrt{2}$ and $\ket{\psi} = (\ket{\mathrm{s}}+\textrm{e}^{\textrm{i}\phi_2}\ket{\mathrm{l}})/\sqrt{2}$, where $\phi_1$ and $\phi_2$ are the relative phases between short and long time-bin paths for either photon, leads to:
%%%%%%%%%%%%%%%%%%%%%%%%%%%%%%%%%%%%%%%
\begin{eqnarray}
\ket{\Upsilon} &=& \frac{1}{\sqrt{2}} \left[ \ket{\Phi(\pi + 2\phi_1)} \cos \frac{\phi}{2} \right.  \nonumber \\ 
&& + \left. \mathrm{e}^{\frac{\mathrm{i}}{2} (\phi+\pi + 2\phi_1)} \ket{\Psi^-} \sin \frac{\phi}{2} \right] \, ,
\label{equ5}
\end{eqnarray}
%%%%%%%%%%%%%%%%%%%%%%%%%%%%%%%%%%%%%%%
in which we have set $\phi_2 = \pi + \phi_1$. By setting $\phi=0$, and changing the two other phases, we have access to any superposition of $\ket{\Phi^-}$ and $\ket{\Phi^+}$, and to these two basis states as well. This simple scheme therefore leads to the possibility of acting on and/or controlling the degree of entanglement. Regarding experimental realizations, note that the heralded photon sources and the entanglement preparation and control circuits could be realized in the near future using integrated optics, as discussed in~\cite{Tanzilli:2011}. This would make our scheme particularly suitable for practical realizations.

In the example above entanglement is generated by particle exchange effects rather than by interactions. The latter degree of entanglement can be quantified either using Bell's inequality or by the von Neumann entropy of the reduced density matrix, after tracing over one of the subsystems (3 or 4). A straightforward calculation of the von Neumann entropy shows that it does depend on the geometric phase. Since the geometric phase is achromatic, we can apply the same phase over all the frequencies in the band of interest by tuning $\varphi_{34}$ and generate entangled photon pairs with the degree of accuracy and the control needed for a source of photonic entanglement.

There have been measurements \cite{Brendel:1995} of the geometric phase in intensity interferometry reported previously, where the geometric phase is also manifest in the lower order correlations, and local in effects at each detector. This may be viewed as two copies of the single photon geometric phase, which either add or cancel (depending on their relative sign) rather than a genuine two particle phase. As a result the phase observed in this experiment is not half the solid angle on the Poincar\'e  sphere, but twice this value or zero. There is also a difference in the nature of the source. The experiment described in \cite{Brendel:1995} uses a pair of entangled photons as a source, while our sources are incoherent and thermal, just like in the original \hbtd experiment.

\section{Conclusion}
\label{conclusion}
To conclude, we have experimentally demonstrated a simple generalization of the \hbt effect, making use of the vector nature of light in order to produce a nonlocal geometric phase. The only conceptual difference between the proposed experiment and the classic \hbt experiment is the presence of polarizers at the sources and detectors. These polarizers cause a geometric phase to appear in the coincidence counts between the two detectors that receive linearly polarized light. Neither the single count rates, nor the self correlations of individual detectors show any geometric phase effects. These appear solely in the \emph{cross} correlation between the count rates of the detectors. This experimental demonstration will open up possibilities for entanglement tuning via the geometric phase.

\acknowledgments
We acknowledge support from the CNRS, the University of Nice - Sophia Antipolis, the Agence Nationale de la Recherche (ANR) for the `e-QUANET' project (grant agreement ANR-09-BLAN-0333-01), the European ICT-2009.8.0 FET Open program for the `QUANTIP' project (grant agreement 244026), and the Conseil Regional PACA, for financial support.

%\bibliography{%

\begin{thebibliography}{10}
\expandafter\ifx\csname url\endcsname\relax\def\url#1{\texttt{#1}}\fi

\bibitem{Berry:1984}
\Name{Berry M.~V.} \REVIEW{Proc. Roy. Soc. Lond. }{392}{1984}{45}.

\bibitem{Shapere:1989}
\Name{Shapere A. \and Wilczek F.} \Book{Geometric Phases in Physics} (World
  Scientific, Singapore) 1989.

\bibitem{Ramaseshan:1986}
\Name{Ramaseshan S. \and Nityananda R.} \REVIEW{Curr. Sci. }{55}{1986}{1225}.

\bibitem{Pancharatnam:1956}
\Name{Pancharatnam S.} \REVIEW{Proc. Indian Acad. Sci. }{44}{1956}{247}.

\bibitem{Samuel:1988}
\Name{Samuel J. \and Bhandari R.} \REVIEW{Phys. Rev. Lett. }{60}{1988}{2339}.

\bibitem{HanburyBrown:1956}
\Name{{Hanbury~Brown} R. \and Twiss R.~Q.} \REVIEW{Nature }{177}{1956}{27}.

\bibitem{Mehta:2010}
\Name{Mehta P., Samuel J. \and Sinha S.} \REVIEW{Phys. Rev. A
  }{82}{2010}{034102}.

\bibitem{Klyshko:1990}
\Name{Klyshko D.~N.} \REVIEW{Phys. Lett. A }{146}{1990}{93}.

\bibitem{Klyshko:1989}
\Name{Klyshko D.~N.} \REVIEW{Phys. Lett. A }{137}{1989}{334}.

\bibitem{Shih:2011}
\Name{Shih Y.} \Book{An Introduction to Quantum Optics : Photon and Biphoton
  Physics} (CRC Press) 2011.

\bibitem{Brendel:1999}
\Name{Brendel J., Gisin N., Tittel W. \and Zbinden H.} \REVIEW{Phys. Rev. Lett.
  }{82}{1999}{2594}.

\bibitem{Hariharan:2003}
\Name{Hariharan P.} \Book{Optical Interferometry} 2nd Edition (Elsevier Science
  (U.S.A.), San Diego) 2003.

\bibitem{Tittel:2001}
\Name{Tittel W. \and Weihs G.} \REVIEW{Quant. Inf. Comp. }{1}{2001}{3}.

\bibitem{Nielsen:2010}
\Name{Nielsen M.~A. \and Chuang I.~L.} \Book{Quantum Computation and Quantum
  Information} 10th Edition (Cambridge University Press, Cambridge) 2010.

\bibitem{Buttiker:1992}
\Name{B\"uttiker M.} \REVIEW{Phys. Rev. Lett. }{68}{1992}{843}.

\bibitem{Aharonov:1959}
\Name{Aharonov Y. \and Bohm D.} \REVIEW{Phys. Rev. }{115}{1959}{485}.

\bibitem{Neder:2007}
\Name{Neder I., Ofek N., Chung Y., Heiblum M., Mahalu D. \and Umansky V.}
  \REVIEW{Nature }{448}{2007}{333}.

\bibitem{Samuelsson:2004}
\Name{Samuelsson P., Sukhorukov E.~V. \and B\"uttiker M.} \REVIEW{Phys. Rev.
  Lett. }{92}{2004}{026805}.

\bibitem{Splettstoesser:2009}
\Name{Splettstoesser J., Moskalets M. \and B\"uttiker M.} \REVIEW{Phys. Rev.
  Lett. }{103}{2009}{076804}.

\bibitem{Ericsson:2005}
\Name{Ericsson M., Achilles D., Barreiro J.~T., Branning D., Peters N.~A. \and
  Kwiat P.~G.} \REVIEW{Phys. Rev. Lett. }{94}{2005}{050401}.

\bibitem{Arecchi:1965}
\Name{Arecchi F.~T.} \REVIEW{Phys. Rev. Lett. }{15}{1965}{912}.

\bibitem{Martienssen:1964}
\Name{Martienssen W. \and Spiller E.} \REVIEW{Am. J. Phys. }{32}{1964}{919}.

\bibitem{Scarcelli:2004}
\Name{Scarcelli G., Valencia A. \and Shih Y.} \REVIEW{EPL (Europhys. Lett.)
  }{68}{2004}{618}.

\bibitem{Loudon:2000}
\Name{Loudon R.} \Book{The Quantum Theory of Light} 3rd Edition (Oxford
  University Press, Oxford) 2000.

\bibitem{Tanzilli:2011}
\Name{Tanzilli S., Martin A., Kaiser F., {De Micheli} M.~P., Alibart O. \and
  Ostrowsky D.~B.} \REVIEW{to appear in Laser \& Photonics Reviews }{}{2011}{}.

\bibitem{Brendel:1995}
\Name{Brendel J., Dultz W. \and Martienssen W.} \REVIEW{Phys. Rev. A
  }{52}{1995}{2551}.

\end{thebibliography}

\end{document}